# Simple approach for high-contrast optical imaging and characterization of graphene-based sheets

*Inhwa Jung,* [†] *Matthew Pelton,* [‡] *Richard Piner,* [†] *Dmitriy A. Dikin,* [†] *Sasha Stankovich,* [†] *Supinda Watcharotone,* [†] *Martina Hausner,* [∝] *and Rodney S. Ruoff*[*,†]

[†]Department of Mechanical Engineering, Northwestern University, Evanston, Illinois 60208,

[‡]Center for Nanoscale Materials, Argonne National Laboratory, Argonne, Illinois 60439,

[∝] Department of Chemistry and Biology, Ryerson University, Toronto, Ontario

*Corresponding author. Tel: (847) 467-6596. Fax: (847) 491-3915. E-mail: r-ruoff@northwestern.edu

A simple optical method is presented for identifying and measuring the effective optical properties of nanometer-thick, graphene-based materials, based on the use of substrates consisting of a thin dielectric layer on silicon. High contrast between the graphene-based materials and the substrate is obtained by choosing appropriate optical properties and thickness of the dielectric layer. The effective refractive index and optical absorption coefficient of graphene oxide, thermally reduced graphene oxide, and graphene are obtained by comparing the predicted and measured contrasts.

Identifying and characterizing a single nanometer-scale layer, or a small number of layers, of materials such as graphite, any of a number of clays, or metal dichalcogenides such as $WS_2$, is challenging, but is critical for the study of such materials.[1,2] Scanning probe microscopy methods, such as atomic force microscopy (AFM), can both identify the presence of such thin sheets and determine their lateral and vertical dimensions,[3] but are somewhat time-consuming; as well, in order to obtain accurate values of height, thereby discriminating between a single layer of material and a bilayer, a relatively small area must be scanned. Scanning electron microscopy can also, in principle, be used for identification of individual layers *versus* multilayer sheets, but this imaging typically induces the formation of a layer of contaminant in the exposed region.[4]

Optical methods, on the other hand, offer the potential for rapid, non-destructive characterization of large-area samples. Ellipsometry, for example, is widely used to determine the optical constants and thicknesses of thin films. Standard ellipsometers, though, require samples with lateral dimensions well over a millimeter. By contrast, imaging ellipsometry can have sub-micrometer resolution, and may be useful for probing optical constants and thicknesses.[5,6] Investigations into this method are ongoing, and will be reported elsewhere. For the past two years, we have focused on simpler methods that allow the use of standard confocal microscopy for rapid identification and characterization of the optical response of thin sheets.[7,8]

In particular, we have investigated the use of substrates designed to interferometrically enhance the visibility of thin sheets. Interference techniques have been used for over half a century to allow for the imaging of low-contrast and transparent samples.[9,10] Over the past decade, microscopy of fluorescent monolayers has been enhanced by incorporating a thin dielectric layer between the material and a reflective substrate.[11] Fabry-Perot interference in the dielectric layer modulates the fluorescence intensity, allowing the determination of the thicknesses of surface layers with nanometer precision.[12] Recently, a similar method has been used for the identification of single graphene sheets.[1,2] Graphene monolayers and multilayers were deposited on substrates consisting of a silicon wafer with an intermediate, 300-nm thick silicon dioxide layer, and the monolayers were qualitatively identified by their weak contrast under white-light illumination. The contrast between the graphene layers and the substrate has been modeled using a multilayer interference method.[13,14] It was thereby shown that contrast could be improved by using narrow-band illumination, thereby allowing for the straightforward determination of the number of graphene layers.[15,16] Spectral resolution of the contrast was used for approximate determination of the optical properties of graphene layers.[17] However, limited contrast was observed for single layers, with the reflectivity from the thin sheet at most 10% less than the reflectivity from the bare substrate. Here, we show that the substrate can be optimized to provide for much higher-contrast imaging, allowing reflectivity from single layers to be more than 12 times that from the bare substrate. The high visibility allows for systematic comparison of observed contrast to calculations, leading to a quantitative determination of the effective optical properties of the thin sheets on the substrates.

The thin sheets that we focus on here are graphite oxide layers. Graphite oxide is a layered material that can readily be exfoliated to form stable colloidal suspensions in water.[18,19] At an appropriate concentration, evaporation of droplets of such a colloidal suspension on a surface yields almost exclusively individual layers, hereafter referred to as graphene oxide. Graphene oxide is not electrically conductive, according to our own measurements, and its optical properties are thus expected to differ markedly from those of graphite. We have found that the electrical conductivity of individual layers of graphene oxide can be increased by heating them in vacuum or through appropriate chemical reduction, making the material of interest in electrical devices.[20] As we show here, the thermal treatment of the graphene oxide layers also alters their effective optical properties.

In order to optimize the substrate, we have calculated the expected contrast between light reflected from the sample with and without the graphene oxide layer. Figure 1-(a) shows a schematic of optical reflection and transmission for the layered thin-film system. For a dielectric film on a substrate (here, Si), the system has two interfaces; with the addition of the graphene oxide layer, a third interface is added. Because a portion of the beam is reflected from each interface and the rest is transmitted, an infinite number of optical paths are possible. The amplitude of the reflected beam is a result of interference between all these paths, and is determined by the wavelength of the incident beam, $\lambda$; its incident angle, $\theta$; the refractive indices, *n*, of the layers; their absorption coefficients, *k*; and the layer thicknesses, *d*.[21,22] In this model, we treat the graphene-oxide sheet as a thin slab of material with a certain refractive index and absorption coefficient. These properties are expected to be different from an isolated sheet of material in vacuum, and from those of bulk graphene oxide, due to the possible presence of adsorbates and local-field effects.

The total amplitude of reflected light depends on the phase changes across the thin material ($\delta_2 = d_2(n_2 - ik_2)\cos\theta_2 2\pi/\lambda$) and the dielectric layer ($\delta_3 = d_3(n_3 - ik_3)\cos\theta_3 2\pi/\lambda$), as well as the amplitudes of the light reflected at the three interfaces, between air and the thin layer of material ($r_1$), between the thin layer of material and the dielectric layer ($r_2$), and between the dielectric layer and the silicon layer ($r_3$):

$$r_{platelet} = \frac{r_1 + r_2\exp(-2i\delta_2) + [r_1 r_2 + \exp(-2i\delta_2)]r_3\exp(-2i\delta_3)}{1 + r_1 r_2\exp(-2i\delta_2) + r_3\exp(-2i\delta_3)[r_2 + r_1\exp(-2i\delta_2)]} \quad (1)$$

In the absence of the thin layer, the total reflected amplitude is expressed in terms of the phase change across the dielectric layer ($\delta_2' = d_3(n_3 - ik_3)\cos\theta_2' 2\pi/\lambda$) and the amplitudes of the light reflected at the two interfaces, between air and the dielectric layer ($r_1'$) and between the dielectric layer and the silicon layer ($r_2'$):

$$r_{dielectric} = \frac{r_1' + r_2'\exp(-2i\delta_2')}{1 + r_1' r_2'\exp(-2i\delta_2')} \quad (2)$$

The reflected intensity can be obtained by multiplying the reflected beam amplitude by its complex conjugate.[23] The reflected intensity is averaged over the angle of incidence, $\theta$, assuming that the incident laser beam has a Gaussian intensity profile. The integration was taken from normal incidence to the maximum incident angle, determined by the numerical aperture of the microscope objective and the diameter of the incident beam. For calculation of the reflection amplitudes and phase changes and details on the angular integration, see the Supporting Information.

The visibility of the graphene oxide films is characterized in terms of the Michelson contrast:[24]

$$contrast = \frac{R_{material} - R_{dielectric}}{R_{material} + R_{dielectric}}, \quad (3)$$

where $R_{material}$ is the reflected intensity with the material and $R_{dielectric}$ is the intensity without the material. If the value of the contrast is zero, the material is not detectable; if the value is between 0 and -1, the material appears darker than the substrate; and if it is between 0 and +1, the material is brighter than the substrate.

Silicon dioxide and silicon nitride are commonly deposited on silicon, and are thus good candidate dielectrics for enhancing the contrast of thin layers. Their suitability can be determined by calculating the expected contrast as a function of the incident wavelength and of the dielectric film thickness. We consider silicon dioxide deposited by a wet thermal oxidation method and silicon nitride deposited by a stoichiometric growth method, as described in the Supporting Information. Figure 1-(b) shows the measured indices of refraction for these two films. Both dielectrics have negligible extinction coefficient. Based on these optical properties, the contrast was calculated, assuming, as a starting point, that the graphene oxide layer has $n = 2$, $k = 0$, and $d = 1$ nm. Results for the silicon dioxide layer are shown in figure 1-(c-1, c-2). The contrast oscillates slightly around zero, with a period that increases with the wavelength of the incident light. Calculated contrast for the silicon-nitride intermediate layer is shown in figure 1-(d-1, d-2). Compared to the silicon–dioxide case, the contrast oscillates at a higher frequency as thickness increases, with significantly higher maxima and significantly lower minima.

With a silicon nitride substrate, high contrast can be obtained for a wide range of optical properties of the thin sheets. Figure 1-(e, f) shows the change in the contrast as the refractive index and absorption coefficient of the thin layer are changed, for optimized dielectric-layer thickness and optical wavelength. These results imply that the use of silicon-nitride intermediate layers will be effective not only for detecting thin sheets, but also for fitting their effective optical properties, provided that a narrow-band light source, such as a laser, is used for illumination.

In order to verify this prediction, we prepared silicon wafers coated with 61.2, 63.0, 65.5, 67.7, and 71.9 nm of silicon nitride. The thicknesses were measured by spectroscopic ellipsometry, and were found to have a variation of approximately 9% across a four-inch wafer. Figure 2 shows confocal microscope images of a single layer of graphene oxide deposited on the 72-nm thick silicon nitride substrate. As expected, the contrast changes depending on the wavelength of the incident light. The normalized intensity profiles of a graphene oxide sheet, measured at the same cross-section before and after thermal treatment, clearly differ, as shown in figure 2-(c), (g), (k). When the sheet thickness was checked by scanning the same region by AFM, no significant change of the thickness of the material was detected, implyinh that the effective optical properties of the material have been changed. Figure 2-(e), (h), (l) show the calculated contrast, as defined in Eq. (1), as a function of the effective optical properties of the material layer. Two lines of contrast, before and after thermal treatment, are drawn on the surface and are projected on the plane of optical properties of the material layer. (Details of how experimental contrast is determined are given in the Supporting Information.) Although the lines clearly show that the effective optical properties have changed due to the thermal treatment, they do not provide sufficient information to give values of $n$ and $k$. We therefore systematically investigated the dependence of the contrast on graphene-oxide layer thickness, dielectric layer thickness, and optical wavelength, and fit the results to theoretical calculations.

Figure 3 shows images of graphene oxide sheets with folded structures on the 64-nm thick silicon nitride layer. The number of graphene-oxide layers can be determined by evaluating the folds in the sheets. Up to four layers of a folded graphene oxide sheet can be identified in the section AA-AA'. For an incident wavelength of 543 nm, very large contrast changes result from thermal treatment of the material. Since the contrast change is related to the effective optical properties and the thickness of the material layer, the thickness was measured by AFM before and after the thermal treatment. The results, given in Table I, show that the thicknesses of multiple layers of graphene oxide are reduced after the thermal treatment with the decrement becoming larger as the number of layers increases. This suggests a decrease in the thickness of intercalated water layers,[25] an important issue for studying multilayer stacks of graphene oxide.[26,27]

In figure 4-(a), (d), (g), the measured and calculated contrasts are shown as a function of the measured graphene oxide thickness. The best fit was found by minimizing the mean square error between the measured and calculated contrast:

$$mean\ square\ error = \sum_{i=1}^{N} (measured\ contrast_i - calculated\ contrast_i)^2 / N \, , \qquad (4)$$

where $N$ is the number of data points. The adjustable parameters are the optical constants ($n$ and $k$) of the material layer; the thickness of the material was assumed to be 1 nm (corresponding to thicknesses measured by AFM). In figure 4-(b), (c), (e), (f), (h), (i), the mean square error is shown on the plane of optical properties of the material layer. The mean square errors for incident wavelengths of 488 nm and 633 nm do not have a single minimum in the plane of optical properties, but the 543-nm result has a localized minimum, clearly showing that both the effective extinction coefficient and the index of refraction increased from the thermal treatment. We thus suggest that the thermal treatment has partially reduced the graphene oxide sheet, increasing both $n$ and $k$ towards the values for pristine graphene.

In figure 5-(a), (d), (g), contrasts measured from substrates with different silicon-nitride thicknesses are shown, along with calculated values. The fitting results again show that the effective index of refraction and extinction coefficient both increased as a result of the thermal treatment (see figure 5-(e), (f)).

We also investigated the effect on the contrast of changing the optical wavelength, for a fixed dielectric-layer thickness of 67 nm, by performing confocal microscope measurements with a continuously tunable light source. In figure 6-(a), the measured and calculated contrast values are shown at different wavelengths. In this case, we have observed contrasts higher than 0.8, which correspond to reflection from the graphene-oxide layer more than 12 times higher than from the bare substrate. To fit the measured data, the effective optical properties are assumed to obey Cauchy functions:

$n = A_n + B_n / \lambda^2$ and $k = A_k + B_k / \lambda^2$, with constants $B_n$ and $B_k$ taken to be 3000 and 1500, respectively.[28] The results indicate that that $A_n$ increases from 2.0 ±0.2 to 2.1 ±0.2 and $A_k$ from 0.3 ±0.1 to 0.6±0.1. With the aforementioned values for $B_n$ and $B_k$, the values of *n* before and after thermal treatment are 2.0 and 2.1, and the values of *k* are 0.30 and 0.55, respectively, all at a wavelength of 633 nm. In order to show that these optical changes are related to physical changes in the material, we have performed transport measurements on the single graphene-oxide sheets, before and after thermal treatment. In figure 6-(b), it was found that the untreated material was electrically insulating, and that it became conductive after thermal treatment.

We also performed preliminary investigations of the effective optical properties of graphene sheets on the same substrates. The thickness of graphene layers were measured in several locations by AFM, and confocal microscope images were acquired for excitation wavelengths of 488, 543, and 633 nm. The relation between the contrast and thickness of deposited graphene was thus obtained, as shown in figure 7-(a), (c), (e). In this case, measurement errors are primarily associated with the uncertainty of the thicknesses determined by AFM. In figure 7-(b), (d), (f), the mean-squared error is shown; it is minimum for $n \sim 2.5$ and $k \sim 1.3$. These values are significantly higher than those for the thermally treated graphene oxide sheet, but lower than the values reported in the literature for bulk graphite.[29]

In conclusion, it has been shown that a single layer of graphene oxide can readily be identified with optical microscopy by use of a properly designed substrate. The contrast between the sheet and the bare substrate depends on the thickness and optical properties of an intermediate dielectric layer, as was confirmed by confocal microscope measurements. We compared the measured contrast with detailed calculations, for graphene oxide layers before and after heating in vacuum. The results suggest that, as a consequence of the thermal treatment, both the effective index of refraction and the effective extinction coefficient increase. Further optimization of the contrast should be possible, for example by using narrower-bandwidth light sources, leading to more precise determination of effective optical properties. The method we have demonstrated should be applicable to thin layers of a wide variety of materials, allowing for their rapid imaging and optical characterization.

**Acknowledgement** We gratefully acknowledge support from the National Science Foundation (CMS-0510212), the Naval Research Laboratory (No. N00173-04-2-C003), and the DARPA Center on Nanoscale Science and Technology for Integrated Micro/Nano-Electromechanical Transducers (iMINT) (Award No: HR0011-06-1-0048). We appreciate assistance by W. Russin (confocal microscopy) and M. Takita (deposition of graphene platelets onto silicon nitride substrates). Work at the Center for Nanoscale Materials was supported by the U. S. Department of Energy, Office of Science, Office of Basic Energy Sciences, under Contract No. DE-AC02-06CH11357. We thank D. Gosztola for technical assistance with variable-wavelength confocal microscopy.

**Supporting Information Available:** Experimental procedure, calculation of the contrast, comparison of dielectric layers and uncertainty of optimized optical properties. This material is available free of charge via the Internet at http://pubs.acs.org.

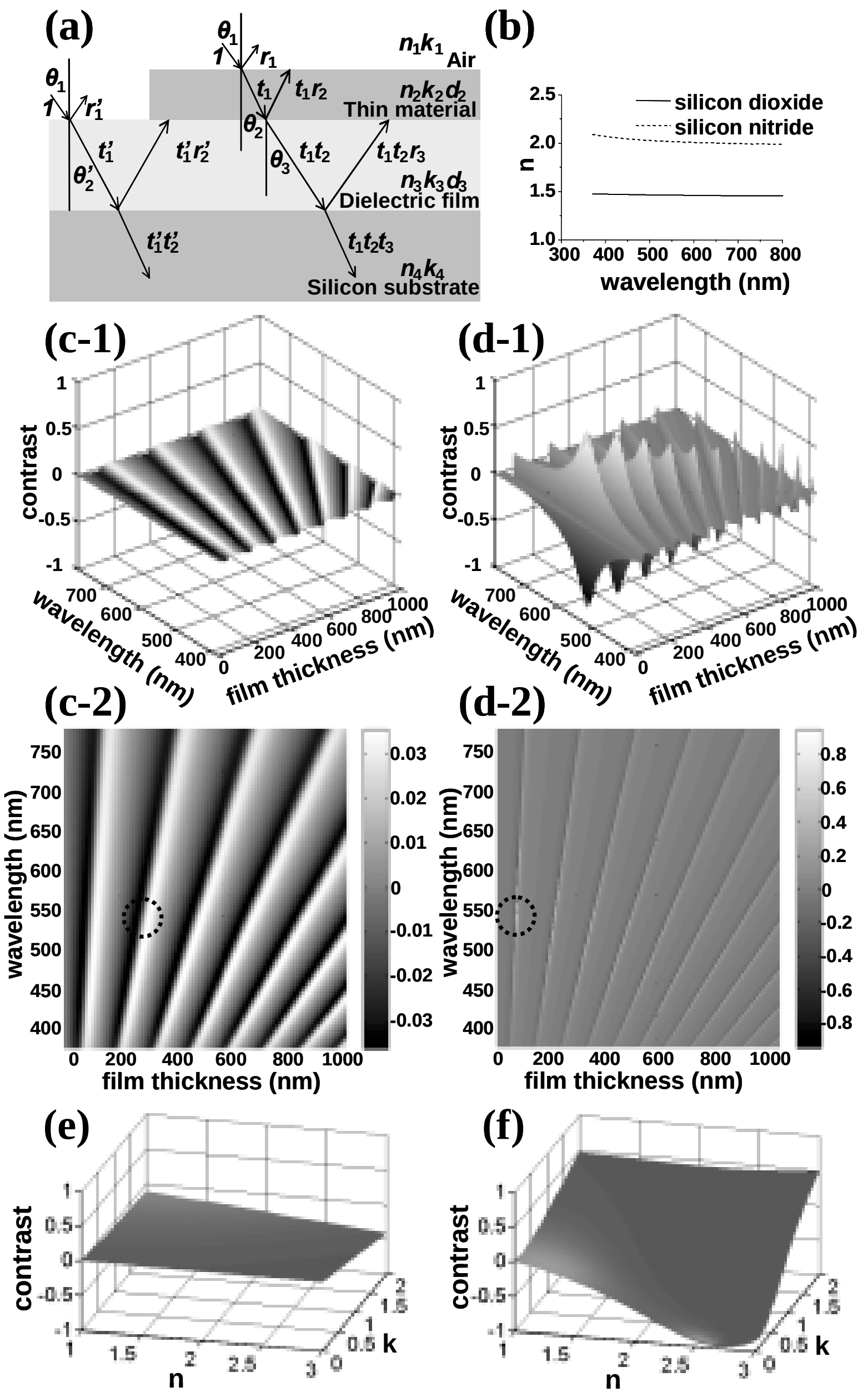

**Figure 1.** (a) Optical reflection and transmission for layered thin-film system: dielectric film on silicon substrate (left), thin sheet added on dielectric film (right). (b) Index of refraction of silicon dioxide (solid line) and stoichiometric silicon nitride (dashed line). (c, d) Contrast as a function of wavelength and film thickness for (c) silicon dioxide layer and (d) silicon nitride layer, each on Si. Calculated contrast dependence on the material refractive index, *n*, and optical absorption coefficient, *k*, for (e) a 275-nm thick silicon dioxide layer, and (f) a 64.5-nm thick silicon nitride layer, assuming an optical wavelength of 543 nm and a numerical aperture of illumination of 0.29.

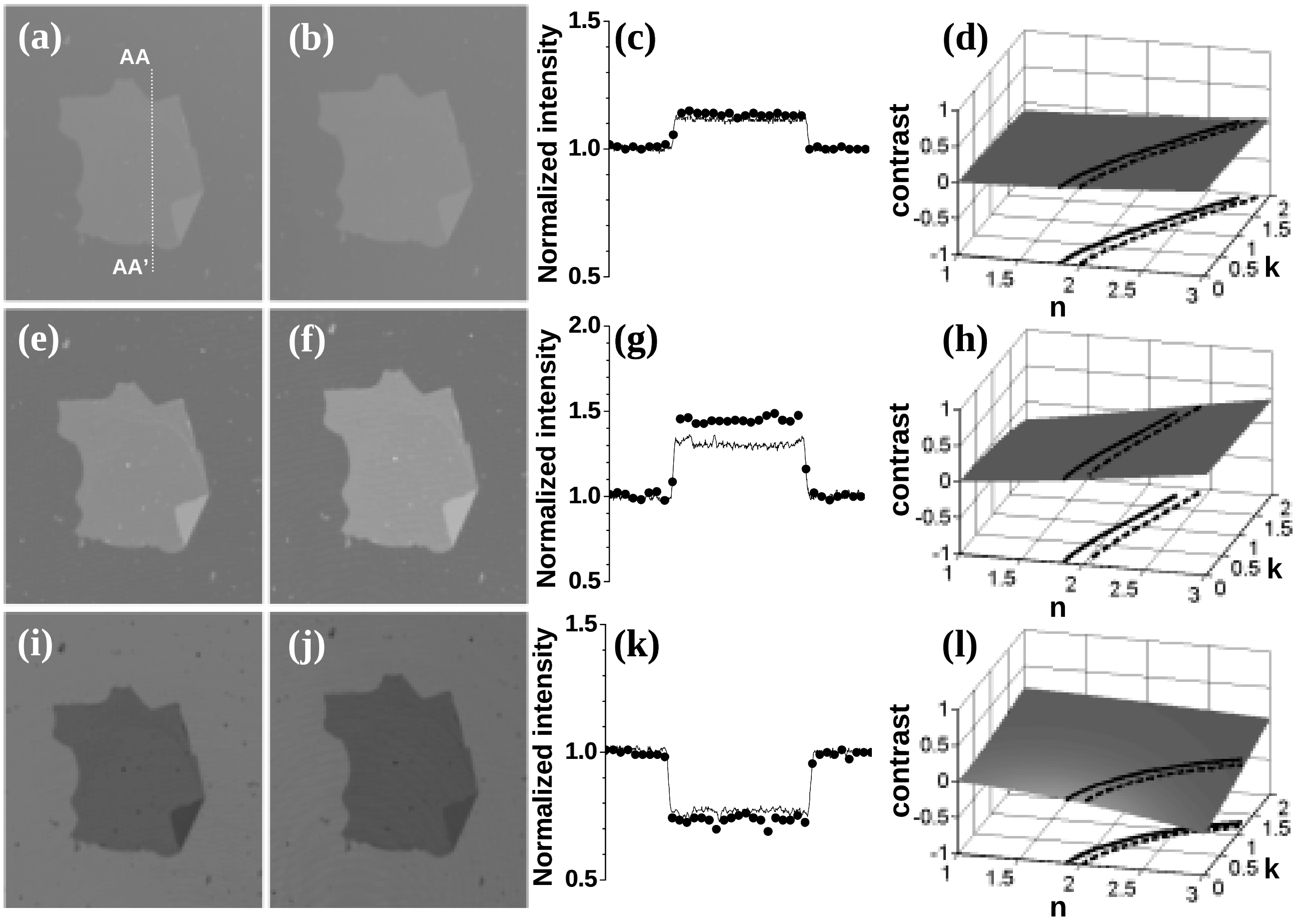

**Figure 2.** Confocal-microscope images of a graphene oxide single sheet with a 72-nm thick silicon-nitride intermediate layer, at three different excitation wavelengths: 488 nm (a, b), 543 nm (e, f), and 633 nm (i, j), before (a, e, i) and after (b, f, j) thermal treatment; Intensity plot across the 100-μm long section AA-AA' before (-) and after (●) thermal treatment: 488 nm (c), 543 nm (g), 633 nm (k); Calculated contrast as a function of refractive index, *n*, and extinction coefficient, *k*, of the graphene oxide sheets at three different wavelengths: 488 nm (d), 543 nm (h), and 633 nm (l). Contrast values measured from confocal images are included to show the change in optical properties before (solid line) and after (dashed line) thermal treatment.

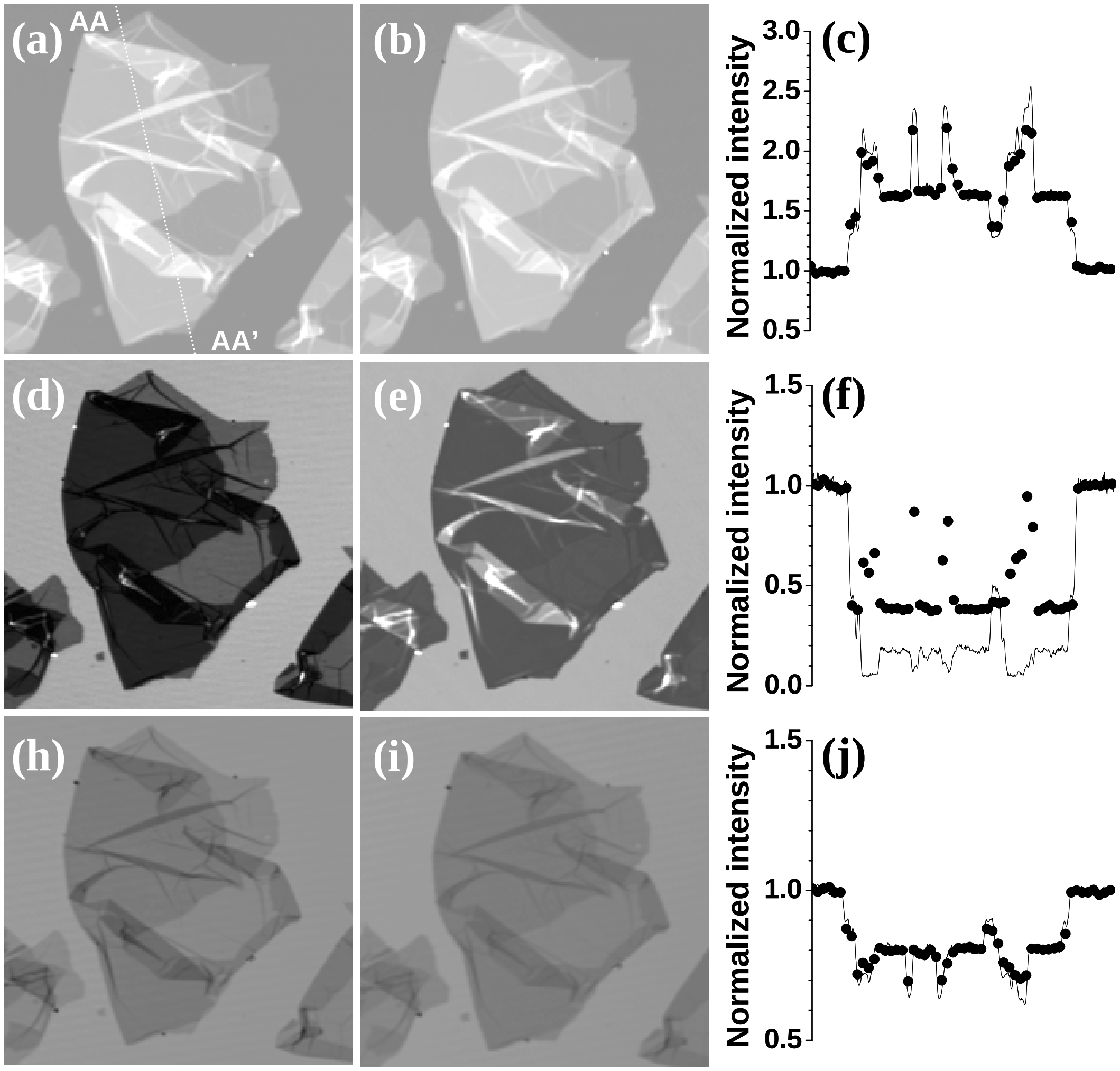

**Figure 3.** Confocal microscope images of a folded graphene oxide sheet with a 64-nm thick silicon-nitride intermediate layer, at three different wavelengths: (a, b) 488 nm, (d, e) 543 nm, and (g, h) 633 nm, before (a, d, g) and after (b, e, h) thermal treatment; Intensity plot across the 125-μm long section AA-AA' before (-) and after (●) thermal treatment, for wavelengths of (c) 488 nm, (f) 543 nm, and (i) 633 nm.

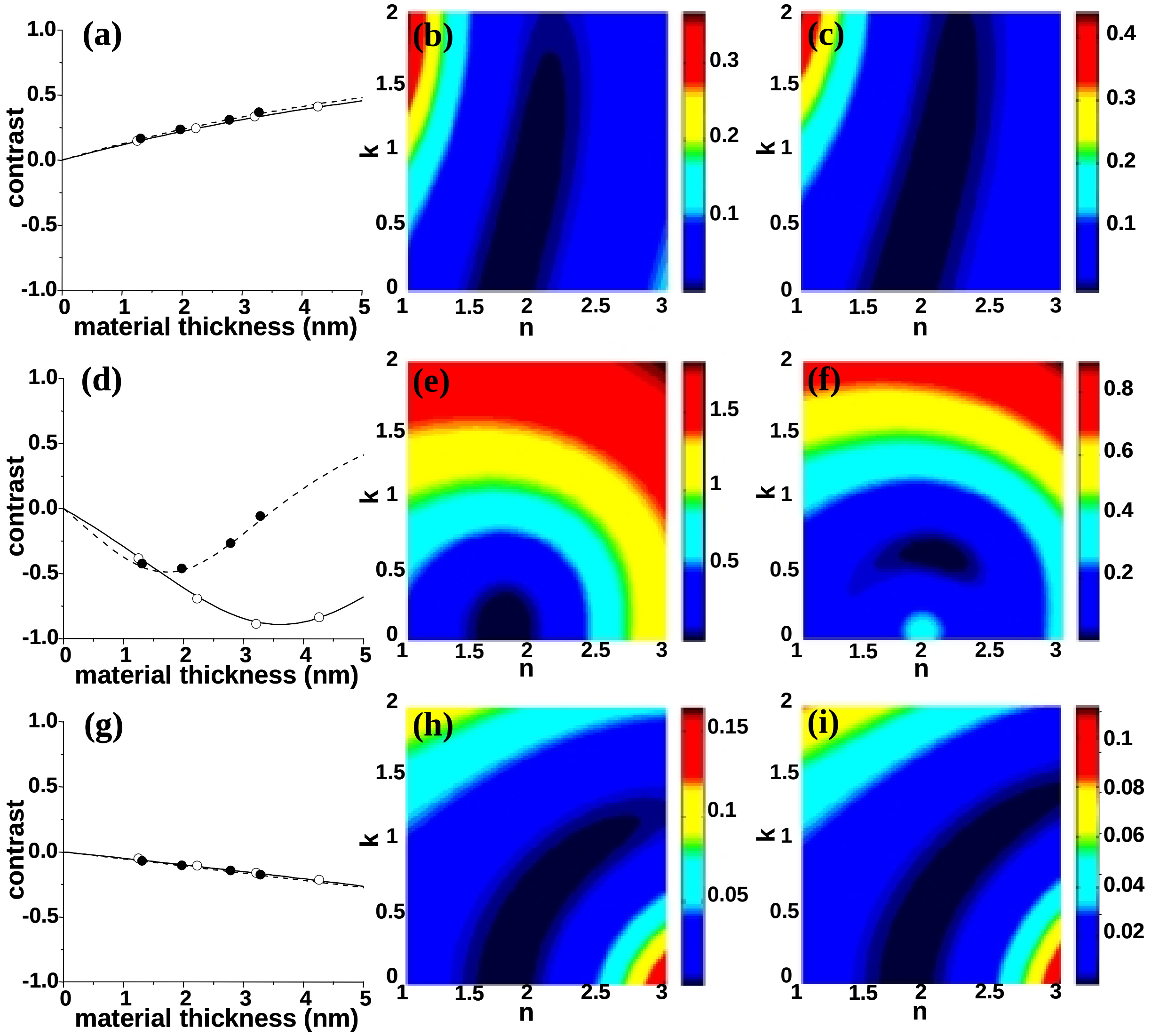

**Figure 4.** Measured contrast *vs* thickness of graphene oxide, before (o) and after (●) thermal treatment, and calculated contrast before (solid line) and after (dashed line) thermal treatment, for wavelengths of 488 nm (a), 543 nm (d), and 633 nm (g); Contour map of the mean square error between the measured and the calculated contrast, for wavelengths of (b, c) 488 nm, (e, f) 543 nm, and (h, i) 633 nm; before (b, e, h) and after (c, f, i) thermal treatment.

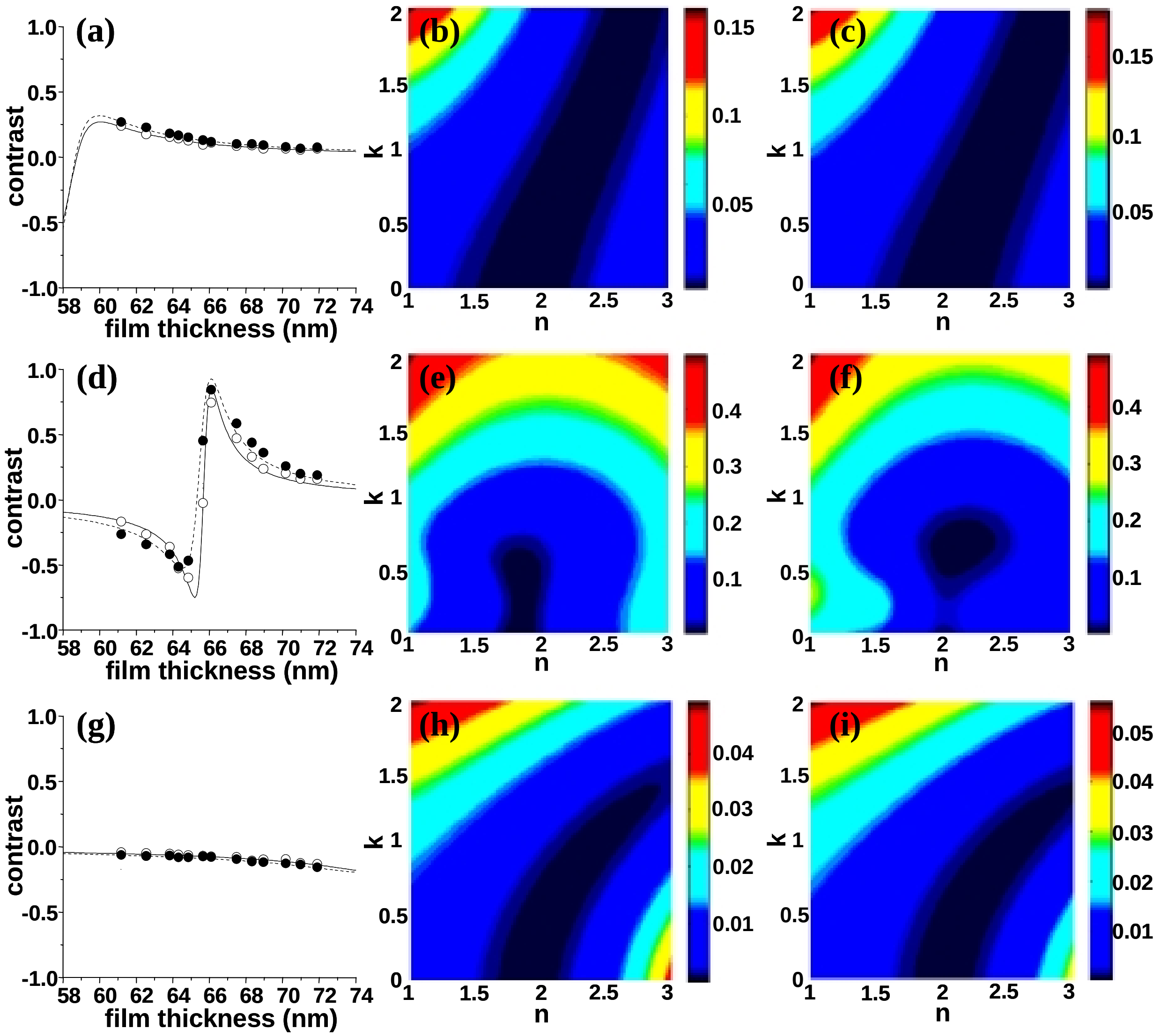

**Figure 5.** Measured contrast *vs.* thickness of intermediate silicon nitride film, before (o) and after (●) thermal treatment, and calculated contrast before (solid line) and after (dashed line) thermal treatment, at wavelengths of 488 nm (a), 543 nm (d), and 633 nm (g); Contour map of the mean square error between the measured and the calculated contrast for wavelengths of (b, c) 488 nm, (e, f) 543 nm, and (h, i) 633 nm, before (b, e, h) and after (c, f, i) thermal treatment.

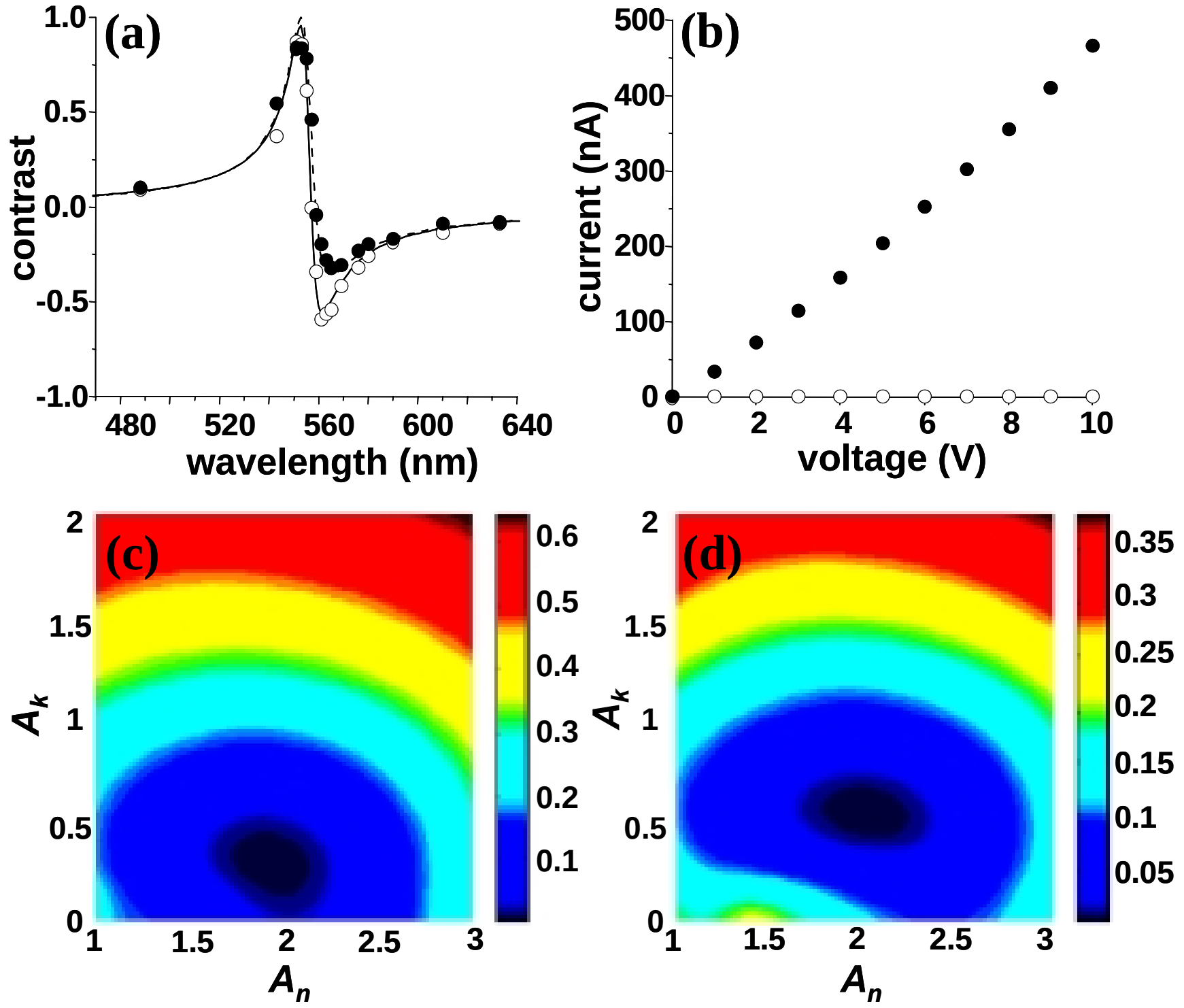

**Figure 6.** Measured contrast *vs*. wavelength of incident light before (o) and after (●) thermal treatment, and calculated contrast before (solid line) and after (dashed line) thermal treatment (a). Current *vs*. voltage for a single graphene oxide sheet before (o) and after (●) thermal treatment (b). Contour map of the mean square error between the measured and the calculated contrast before (c) and after (d) thermal treatment. $A_n$ and $A_k$ are coefficients of the Cauchy functions, $n = A_n + B_n / \lambda^2$ and $k = A_k + B_k / \lambda^2$. The constant values $B_n$ and $B_k$ are assumed to be 3000 and 1500, respectively.

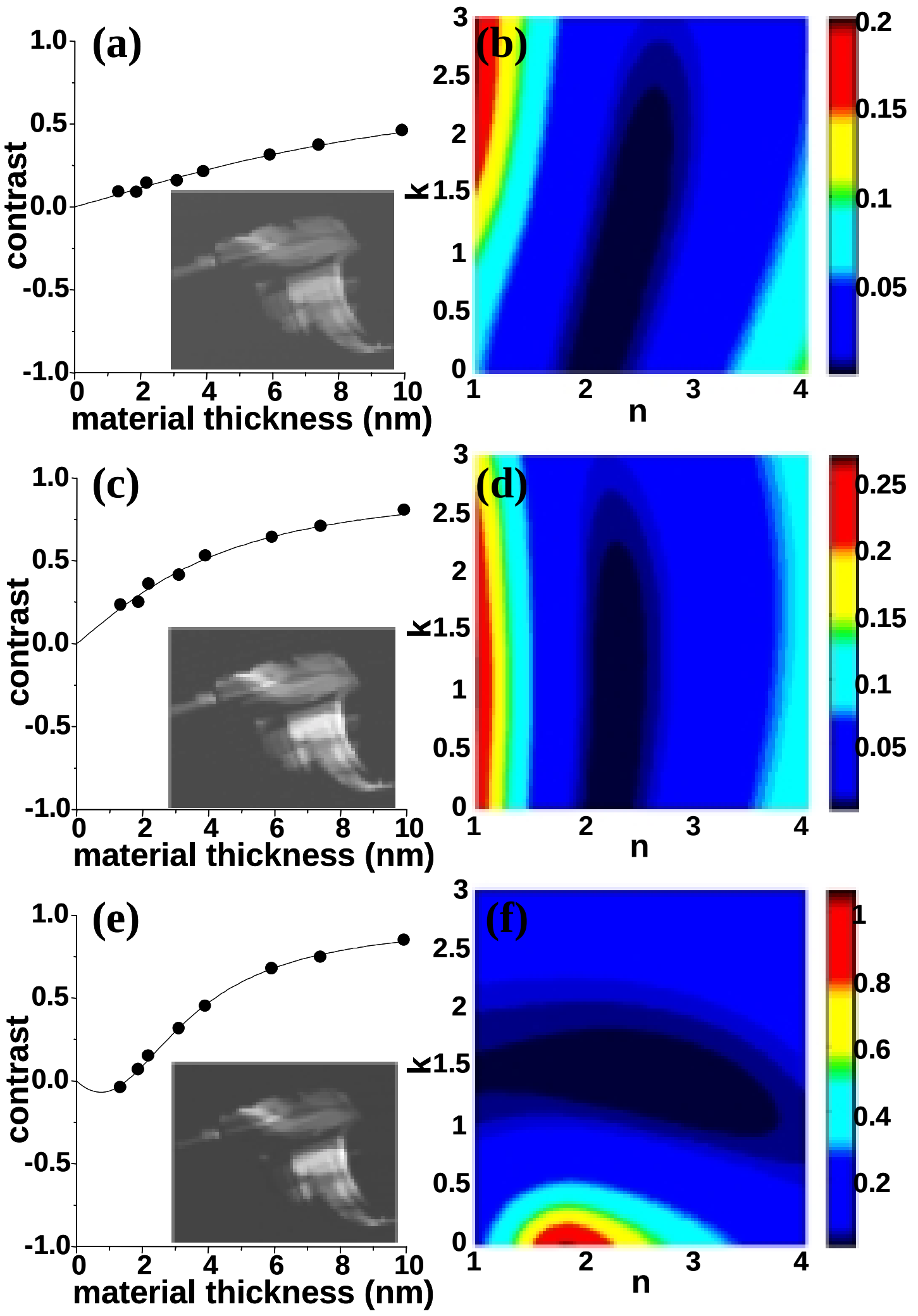

**Figure 7.** Measured contrast *vs*. thickness of graphene (points), and calculated contrast (lines), at three different excitation wavelengths: 488 nm (a), 543 nm (c), and 633 nm (e); Confocal microscope images at each wavelength (insets); Contour map of the mean square error between the measured and the calculated contrast for wavelengths of (b) 488 nm, (d) 543 nm, and (f) 633 nm.

**Table 1.** Layer thicknesses measured by AFM

| Number of layers | Thickness before thermal treatment | Thickness after thermal treatment |
|---|---|---|
| 1 | 1.25 ± 0.08 nm | 1.31 ± 0.10 nm |
| 2 | 2.23 ± 0.11 nm | 1.97 ± 0.08 nm |
| 3 | 3.21 ± 0.21 nm | 2.79 ± 0.14 nm |
| 4 | 4.26 ± 0.25 nm | 3.28 ± 0.06 nm |

## SUPPORTING INFORMATION

### 1. Experimental procedure

#### *General*

The substrate was prepared as follows. First, a dielectric film was grown on a silicon wafer (p type, prime grade, (100) direction, from Helitek), and then alignment marks were patterned on the surface. Graphene oxide sheets were then deposited on the substrate, and alignment marks were used to find the same area with different equipment, including the confocal microscope (Leica LCS laser confocal microscope SP2 system) and AFM (Park Scientific AutoProbe CP/MT scanning probe microscope).

#### *Dielectric film growth*

The silicon dioxide thin film was grown by thermal oxidation at 1100°C (atmospheric oxidation furnace from Bruce Technologies, Inc., located in the Nanotechnology Core Facility at the University of Illinois at Chicago).

Rob Ilic of the Cornell NanoScale Science & Technology Facility grew silicon nitride on silicon with a PECVD system (load-locked GSI Plasma Enhanced Chemical Vapor Deposition system). Two different types of silicon nitride were grown: a "stoichiometric" nitride and a "low-stress" nitride. While the extinction coefficient of the stoichiometric nitride is negligible, the extinction coefficient of the low-stress nitride is larger. The stoichiometric nitride was thus used, since a non-zero extinction coefficient was shown to reduce the contrast. The surface roughness measured by AFM was 0.27 nm; considering that the thickness of the graphene oxide sheets is ~1nm, this value was deemed acceptable for AFM scanning.

Film thicknesses were measured with a spectroscopic ellipsometer (MV-2000; J. A. Woollam, Inc.).

### *Patterning of alignment marks*

Squares 4 μm x 4 μm in size were patterned in 10 x 10 blocks; in order to cover a large area of the substrate, a 10 x 10 array of blocks was patterned. A survey of distances separating the squares (30 μm, 50 μm, and 75 μm) was undertaken to find an inter-square separation that was compatible with the lateral dimensions of the graphene oxide sheets; 50 μm was chosen for subsequent deposition and imaging. To pattern the alignment marks on the substrate, an optical lithography system (MA6; Karl Suss, Inc.) was used. Ti, followed by Au, was then deposited and the masked area was lifted off. 1-nm-thick Ti and 2-nm-thick Au were chosen in order to allow clear imaging by the confocal microscope. The thickness of the metal during deposition was measured by a quartz crystal monitor (XTM/2 Thin Film Deposition Monitor from Inficon) in the deposition system (Varian 3117 E-beam Evaporator).

### *Graphene oxide sheet deposition*

Prior to deposition of the graphene oxide material, which was prepared by the Hummers method,[1] the substrate was sonicated, first in acetone (VWR International, reagent grade) for ~5 min and then in isopropanol (J.T. Baker, reagent grade) for ~3 min; nitrogen gas (Airgas, Inc.) was then blown over the surface to dry the sample. The substrate was then treated with oxygen plasma for 3 min in an oxygen plasma cleaner (Plasma Preen II-862 from Plasmatic systems, Inc.). This process rendered the surface hydrophilic, so that the suspension of graphite oxide wetted the surface well, leading to homogeneous dispersion of the graphene oxide material on the substrate. To minimize the number of graphene oxide sheets deposited from the applied droplet, the suspension was diluted 100 times from is normal value,[2] resulting in a density of 0.01mg/mL. After depositing a droplet on the substrate, the sample was blown with nitrogen gas before drying in air. The time span between deposition of the droplet and blowing with $N_2$ was varied so as to get sparsely dispersed material; based on different trials, a delay of 1 min was used.

***Thermal treatment of graphite oxide***

The deposited graphene oxide sheet material was thermally treated in a vacuum oven (Isotemp vacuum oven Model 280A; Fischer Scientific) attached to a roughing pump (Duo seal vacuum pump from Welch Vacuum Technology, Inc., $1\cdot10^{-4}$ Torr base pressure). The sample was inserted at room temperature, and the oven was ramped up to 200°C in 1 h, held at 200°C for 2 h, and then ramped down to room temperature in 4 h.

***Confocal laser scanning microscope imaging***

Three different laser sources and wavelengths (Ar - 488nm, He:Ne - 543 nm, He:Ne - 633 nm) were used for confocal laser-scanning microscopic imaging. A photomultiplier tube (PMT) was used to detect laser light reflected from the surface. The offset of the PMT was set at zero, and the gain was set to yield a good image contrast. The response of the PMT was assumed to be linear with respect to the light flux within a 3% limit. The beam expander was set to 3X, giving a beam diameter of 9.8mm. The objective used was Leica HC PL APO CS 10X, with numerical aperture of 0.4.

The contrast was determined from the measured images using ImageJ software (Wayne Rasband, National Institutes of Health, http://rsb.info.nih.gov/ij/). Because the background intensity was not even over the entire image, a flattening procedure was applied. As shown in figures 2-(c), (f), (i) and 4-(c), (f), (i), a typical intensity profile can be divided into three sub-profiles: the profile across the sheet, and two background profiles on the left and right sides of the sheet. The background profiles were fit with straight lines, and the normalized intensity was calculated by dividing the intensity profile for the sheet by the fitted lines for the background. The contrast was then calculated according to $(I_{platelet}-1)/(I_{platelet}+1)$, where $I_{platelet}$ is the average of the normalized intensity across the sheet; this is equivalent to Eq. (1), above.

### *Variable-wavelength confocal microscope imaging*

Variable-wavelength confocal microscopy was performed on a home-built microscopy system. Incident light was produced by an optical parametric oscillator pumped by a mode-locked Ti:Sapphire laser (Coherent Mira), which has continuously tunable emission with a bandwidth of approximately 5 nm. This light was coupled into an inverted microscope (Olympus IX71), and focused with an objective (Olympus UPLAN APO 20X, numerical aperture = 0.7) on the sample surface. Reflected light from the sample was collected through the same objective and imaged on an optical fiber, which serves as the confocal aperture. The light coupled into the fiber was detected with a photomultiplier (Hamamatsu H5783), whose gain was adjusted for each image in order to maximize the signal-to-noise ratio while remaining in the linear regime of the detector. Images were obtained by scanning the sample over a 50 X 50 μm region using a piezoelectric-driven flexure stage (Mad City Labs Nano-Bio2).

### *AFM scanning*

AFM was used to obtain the thickness of the graphene oxide sheets. The target sheets could be found by comparing images from the confocal microscope with images from the optical microscope in the AFM. Because the profile of the substrate is not uniform, the thickness was adjusted with regard to the profile of the substrate. The distance between two measured points and the angle of the substrate profile was measured in software (Image Processing and Data Analysis 2.1.15 from TM Microscopes), and the height adjustment was calculated based on these values.

### *Conductivity Measurements*

Conductivity measurements were performed in air by a two-probe technique, with an electrode separation of 10 μm. The thermal treatment for the samples characterized electrically was identical to that for the samples measured optically.

## 2. Calculation of contrast

***Multilayer reflectance model***

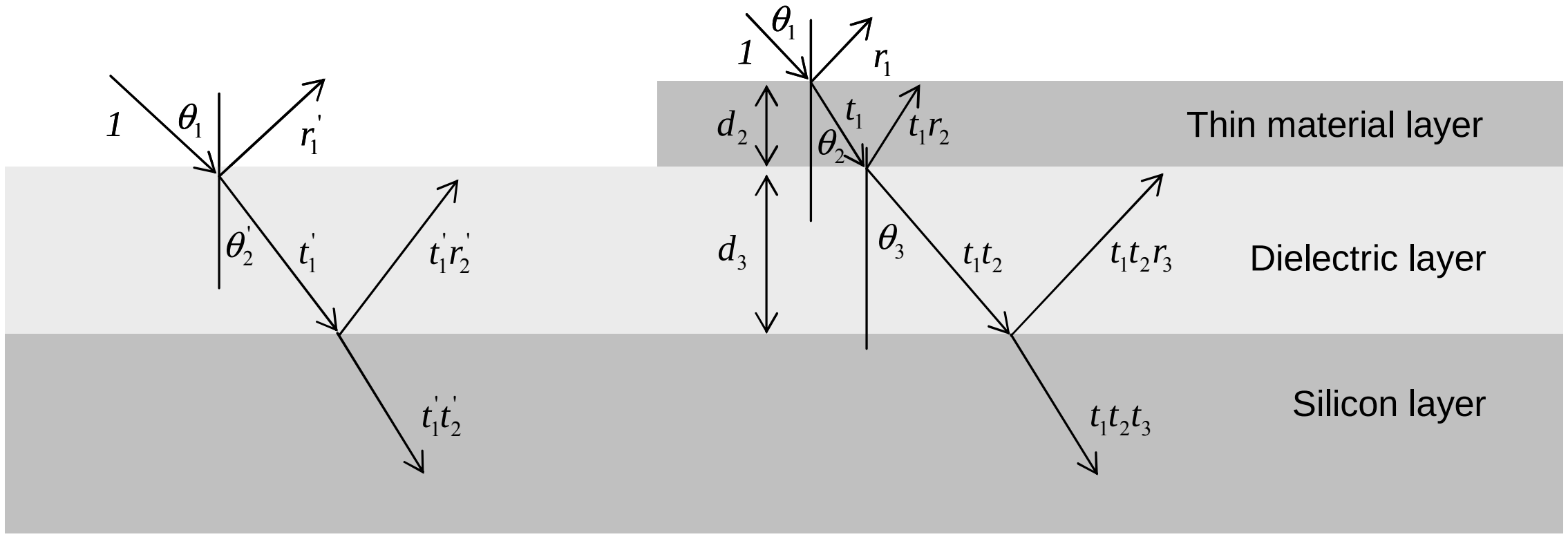

**Figure S-1.** Optical beam reflection and transmission for layered thin film system: dielectric layer on silicon layer (left), thin material added on dielectric layer (right).

As illustrated in figure S-1, the amplitude of light reflected from the thin material ($r_1$) is affected by the thicknesses of the thin material ($d_2$) and the dielectric layer ($d_3$); the optical properties of the thin material ($n_2, k_2$), the dielectric layer ($n_3, k_3$), and the silicon layer ($n_4, k_4$); the wavelength of the incident light ($\lambda$); and the angle of incidence ($\theta_1$). The phase changes across the thin material ($\delta_2$) and the dielectric layer ($\delta_3$) are then

$$\delta_2 = d_2(n_2 - ik_2)\cos\theta_2 2\pi / \lambda \tag{1}$$

$$\delta_3 = d_3(n_3 - ik_3)\cos\theta_3 2\pi / \lambda \tag{2}$$

The total amplitude of reflected light is expressed in terms of these phase changes and the amplitudes of the light reflected at the three interfaces, between air and the thin layer of material ($r_1$), between the thin layer of material and the dielectric layer ($r_2$), and between the dielectric layer and silicon ($r_3$):

$$r_{platelet} = \frac{r_1 + r_2\exp(-2i\delta_2) + [r_1r_2 + \exp(-2i\delta_2)]r_3\exp(-2i\delta_3)}{1 + r_1r_2\exp(-2i\delta_2) + r_3\exp(-2i\delta_3)[r_2 + r_1\exp(-2i\delta_2)]} \tag{3}$$

In the absence of the thin material layer, the phase change and reflected amplitude are

$$\delta_2' = d_3(n_3 - ik_3)\cos\theta_2' 2\pi / \lambda \tag{4}$$

$$r_{dielectric} = \frac{r_1' + r_2'\exp(-2i\delta_2')}{1 + r_1'r_2'\exp(-2i\delta_2')} \tag{5}$$

where $r_1'$ is the reflection amplitude between the dielectric layer and air and $r_2'$ is the reflection amplitude between the dielectric layer and silicon layer. The reflected intensities are

$$R_{platelet} = r_{platelet}\, r^*_{platelet} \tag{6}$$

$$R_{dielectric} = r_{dielectric}\, r^*_{dielectric} \tag{7}$$

The contrast is then defined as follows:

$$contrast = \frac{R_{platelet} - R_{dielectric}}{R_{platelet} + R_{dielectric}} \tag{8}$$

***Calculation of reflection amplitudes***

When the thin material is on the dielectric film, layer 1 is air, layer 2 is the thin material, layer 3 is the dielectric film, and layer 4 is silicon. Equations (1) and (2) can be rewritten as

$$\delta_2 = d_2(u_2 - i\,v_2)\,2\pi/\lambda \tag{9}$$

$$\delta_3 = d_3(u_3 - i\,v_3)\,2\pi/\lambda \tag{10}$$

where $u_2$ and $v_2$ are the real and imaginary parts, respectively, of $(n_2 - ik_2)\cos\theta_2$, and $u_3$ and $v_3$ are the real and imaginary parts, respectively, of $(n_3 - ik_3)\cos\theta_3$. Defining the following quantities:

$$e_{21} = n_2^2 - k_2^2, \quad e_{22} = 2n_2k_2, \quad e_{23} = (n_1\sin(\theta_1))^2 \tag{11-1}$$
$$e_{31} = n_3^2 - k_3^2, \quad e_{32} = 2n_3k_3, \quad e_{33} = (n_1\sin(\theta_1))^2 \tag{11-2}$$
$$e_{41} = n_4^2 - k_4^2, \quad e_{42} = 2n_4k_4, \quad e_{43} = (n_1\sin(\theta_1))^2 \tag{11-3}$$

then *u* and *v* for each of the layers are

$$u_2 = \sqrt{\frac{e_{21} - e_{23} + \sqrt{(e_{21} - e_{23})^2 + e_{22}^2}}{2}}, v_2 = \sqrt{\frac{e_{23} - e_{21} + \sqrt{(e_{21} - e_{23})^2 + e_{22}^2}}{2}} \tag{12-1}$$

$$u_3 = \sqrt{\frac{e_{31} - e_{33} + \sqrt{(e_{31} - e_{33})^2 + e_{32}^2}}{2}}, v_3 = \sqrt{\frac{e_{33} - e_{31} + \sqrt{(e_{31} - e_{33})^2 + e_{32}^2}}{2}} \tag{12-2}$$

$$u_4 = \sqrt{\frac{e_{41} - e_{43} + \sqrt{(e_{41} - e_{43})^2 + e_{42}^2}}{2}}, u_4 = \sqrt{\frac{e_{43} - e_{41} + \sqrt{(e_{41} - e_{43})^2 + e_{42}^2}}{2}}\ . \tag{12-3}$$

The reflection coefficients for *s*-polarization are

$$r_1 = \frac{n_1 \cos(\theta_1) - (u_2 - i v_2)}{n_1 \cos(\theta_1) + (u_2 - i v_2)} \tag{13-1}$$

$$r_2 = \frac{u_2 - i v_2 - (u_3 - i v_3)}{u_2 - i v_2 + (u_3 - i v_3)} \tag{13-2}$$

$$r_3 = \frac{u_3 - i v_3 - (u_4 - i v_4)}{u_3 - i v_3 + (u_4 - i v_4)} \tag{13-3}$$

and the reflection coefficients for *p*-polarization are

$$r_1 = \frac{(n_2^2 - k_2^2)\cos(\theta_1) - n_1 u_2 - i(2n_2k_2\cos(\theta_1) - n_1 v_2)}{(n_2^2 - k_2^2)\cos(\theta_1) + n_1 u_2 - i(2n_2k_2\cos(\theta_1) + n_1 v_2)} \tag{14-1}$$

$$r_2 = \frac{(u_2 - iv_2)(n_3 - ik_3)^2 - (u_3 - iv_3)(n_2 - ik_2)^2}{(u_2 - iv_2)(n_3 - ik_3)^2 + (u_3 - iv_3)(n_2 - ik_2)^2} \tag{14-2}$$

$$r_3 = \frac{(u_3 - iv_3)(n_4 - ik_4)^2 - (u_4 - iv_4)(n_3 - ik_3)^2}{(u_3 - iv_3)(n_4 - ik_4)^2 + (u_4 - iv_4)(n_3 - ik_3)^2} \quad . \tag{14-3}$$

The reflected intensities are averaged over the two polarizations. Similar formulas apply for the two-layer system of dielectric and substrate, when the thin top layer is absent.

***Consideration of incident angle***

The reflected intensity was integrated throughout the angle of incidence, $\theta$, assuming that the intensity profile along the angle of incidence follows a two-dimensional Gaussian distribution. The angular width of the Gaussian is given by $\theta_b = \sin^{-1}(\mathrm{NA}_{effective})$, where the effective numerical aperture of illumination, $\mathrm{NA}_{effective}$, is determined by the numerical aperture, *NA*, of the objective, the back aperture, $d_{aperture}$, of the objective, and the beam diameter, $d_{beam}$: $\mathrm{NA}_{effective} = NA \cdot \left(d_{aperture} / d_{beam}\right)$. The effective numerical apertures for both confocal microscopes are approximately 0.28, as illustrated in Figure S-2. The average was calculated by numerically integrating over ten points between $\theta = 0$ and $\theta = \theta_b$.

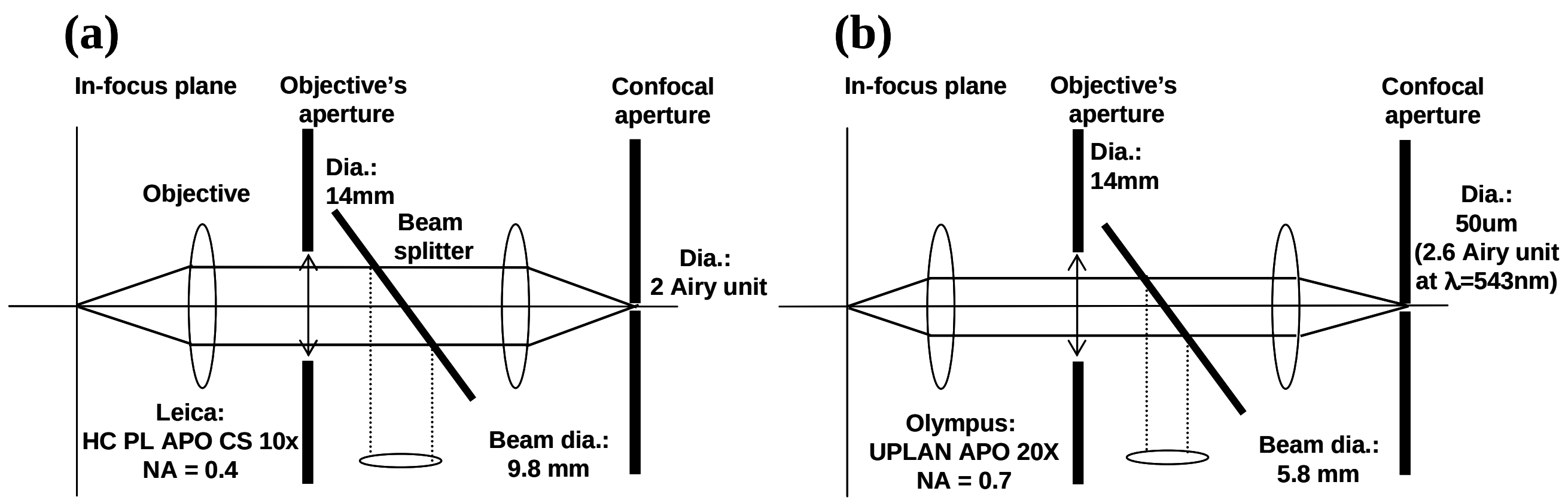

**Figure S-2**. Simplified optical path for confocal microscopy: the fixed-wavelength system (a), the variable-wavelength system (b).

### *Effect of pinhole size on contrast*

The contrast can be strongly affected by the size of the confocal pinhole. However, if the effective numerical aperture of illumination is relatively small, the effect of the pinhole is minimized, as shown in Figure S-3. All measurements reported in the manuscript were taken under these conditions.

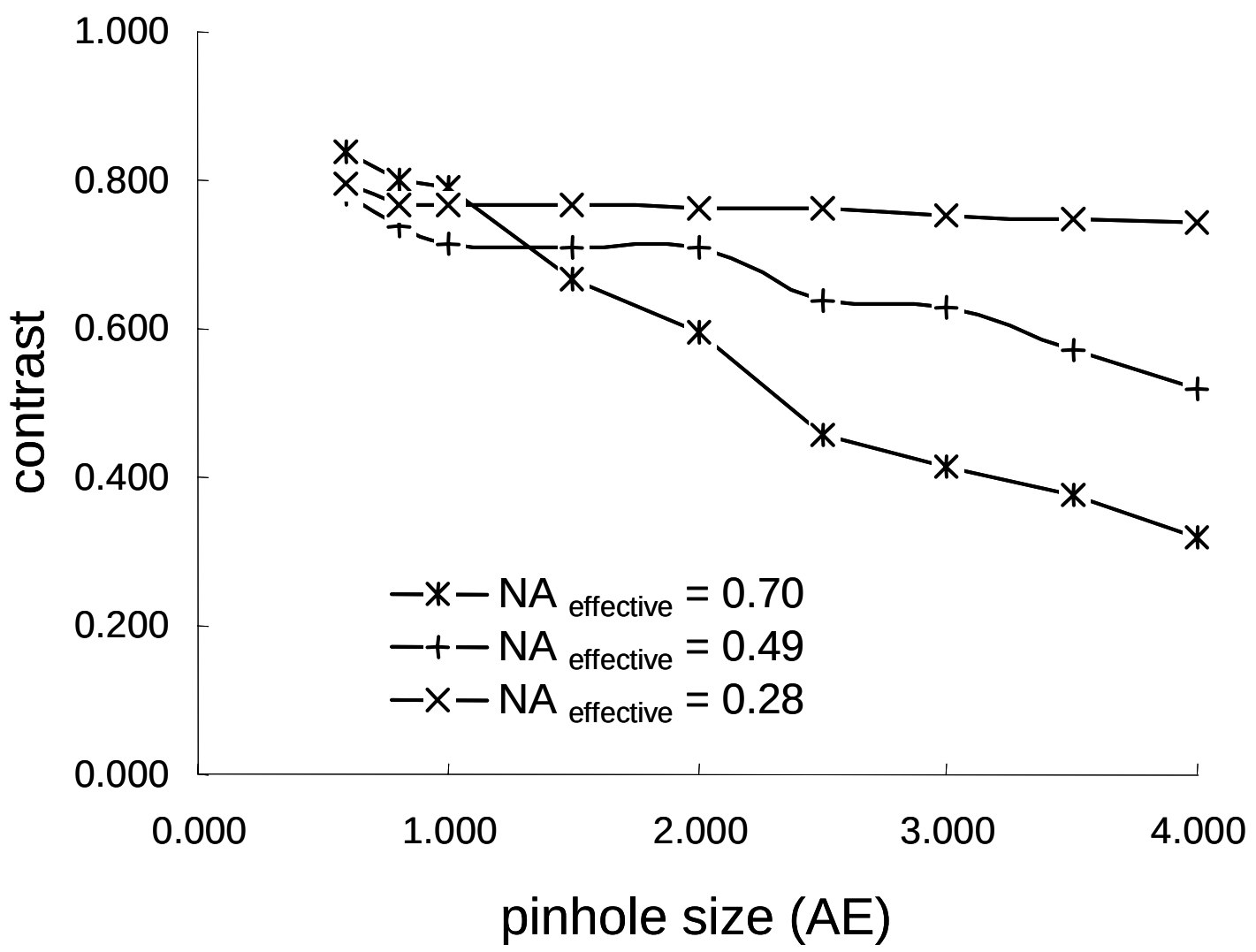

**Figure S-3**. Measured contrast of a single graphene oxide sheet for different pinhole sizes and numerical apertures of illumination. The laser wavelength is 543 nm, and the silicon-nitride thickness is 67 nm.

## *Optical properties of dielectric and substrate materials*

The optical properties of the dielectric films were measured with a spectroscopic ellipsometer. In figure S-4 (a) (b), the measured optical properties of silicon dioxide and silicon nitride are shown. The Cauchy model was used to fit the optical properties and thickness of the as-grown dielectric film. The extinction coefficient is fit as zero for silicon dioxide and 0.0049 for silicon nitride throughout the entire range of wavelengths (380 nm~780 nm). For the optical properties of silicon, data tabulated at the University of Nebraska at Lincoln were used,[3] as shown in figure S-4 (c) (d).

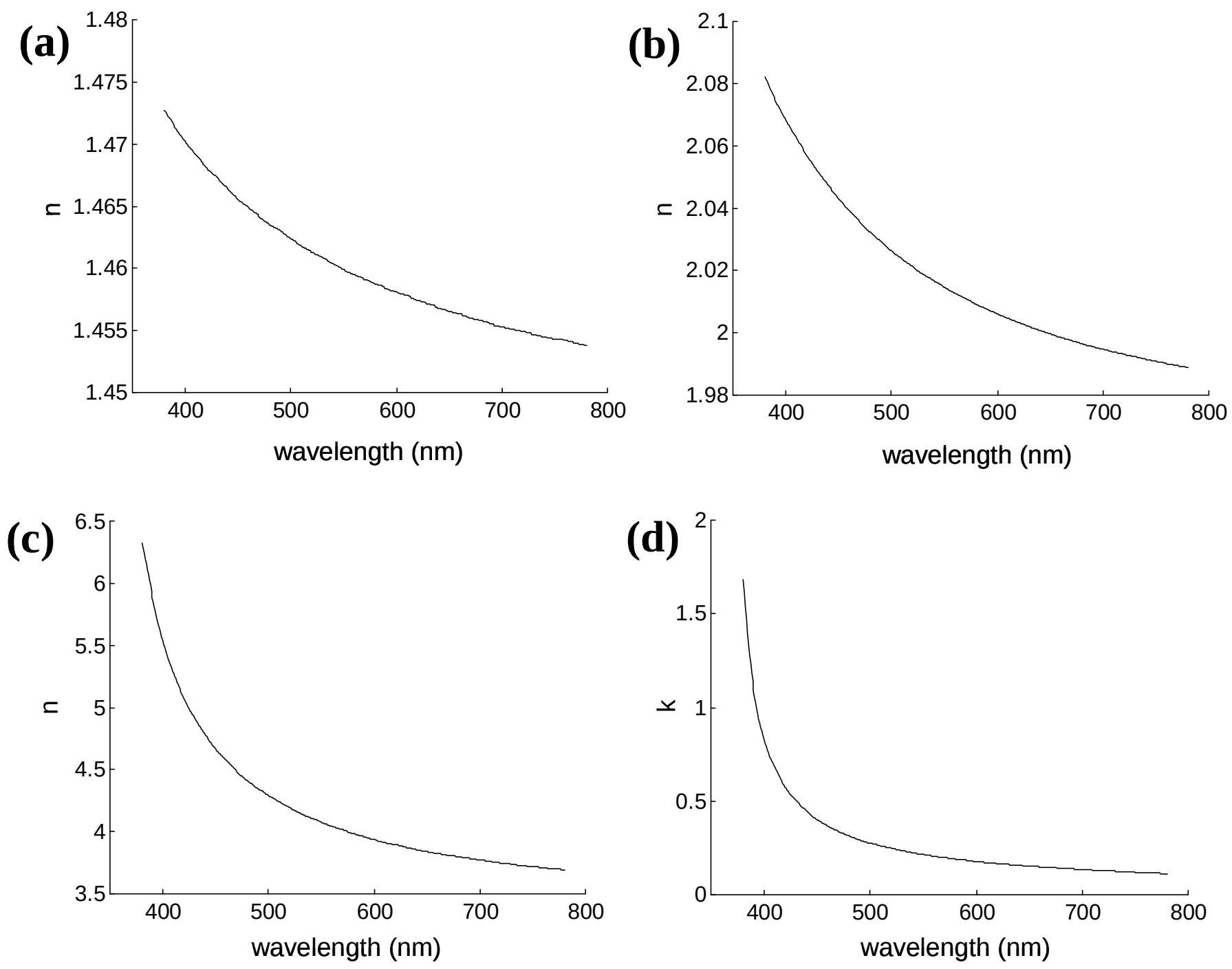

**Figure S-4**. (a) Index of refraction for silicon dioxide and (b) silicon nitride; (c) index of refraction of silicon and (d) its extinction coefficient.

### 3. Comparison of dielectric layers

***Range of thickness for high contrast, and comparison of silicon nitride and silicon dioxide***

As shown in figure S-5, for the common laser wavelengths of 488nm, 543nm, and 633nm, one can expect reasonably high contrast for silicon-nitride layer thicknesses in the range from 55 nm to 80 nm. On the other hand, for silicon-dioxide layers, the contrast does not exceed 0.04.

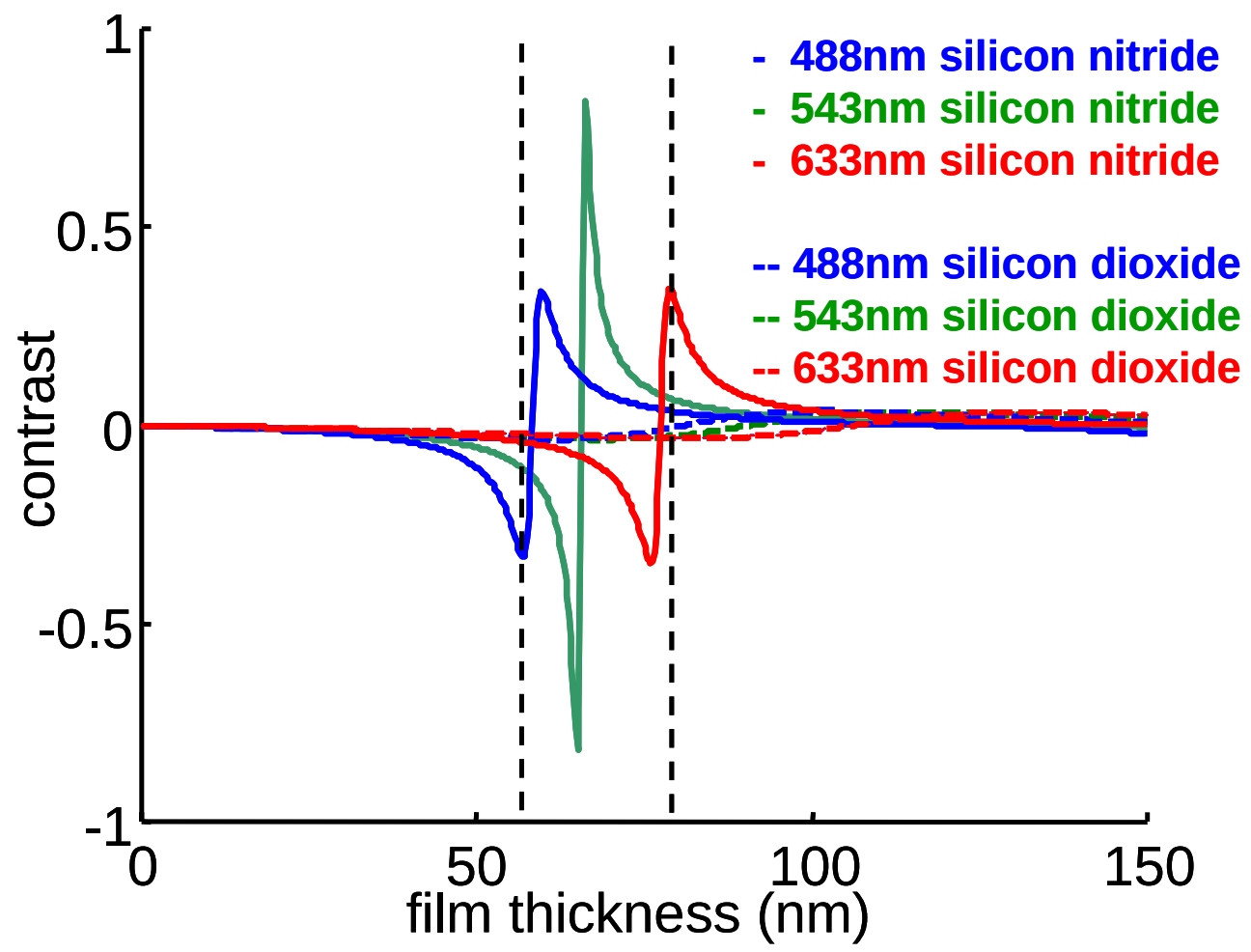

**Figure S-5.** Calculated contrast for thin material with $n$ = 2.0, $k$ = 0, and $d$ = 1.0, for optimized silicon nitride and silicon dioxide layers, and for three incident wavelengths: 488nm, 543nm, and 633nm.

This is illustrated in figure S-6, where confocal-microscope images of graphene oxide single sheets on substrates with silicon-dioxide layers and with silicon nitride layers are compared. The measured contrast values with the silicon dioxide layer were 0.02 at 488nm, -0.02 at 543nm, and -0.01 at 633nm. With the silicon nitride layer, those values were 0.09 at 488nm, 0.68 at 543nm, and -0.07 at 633nm.

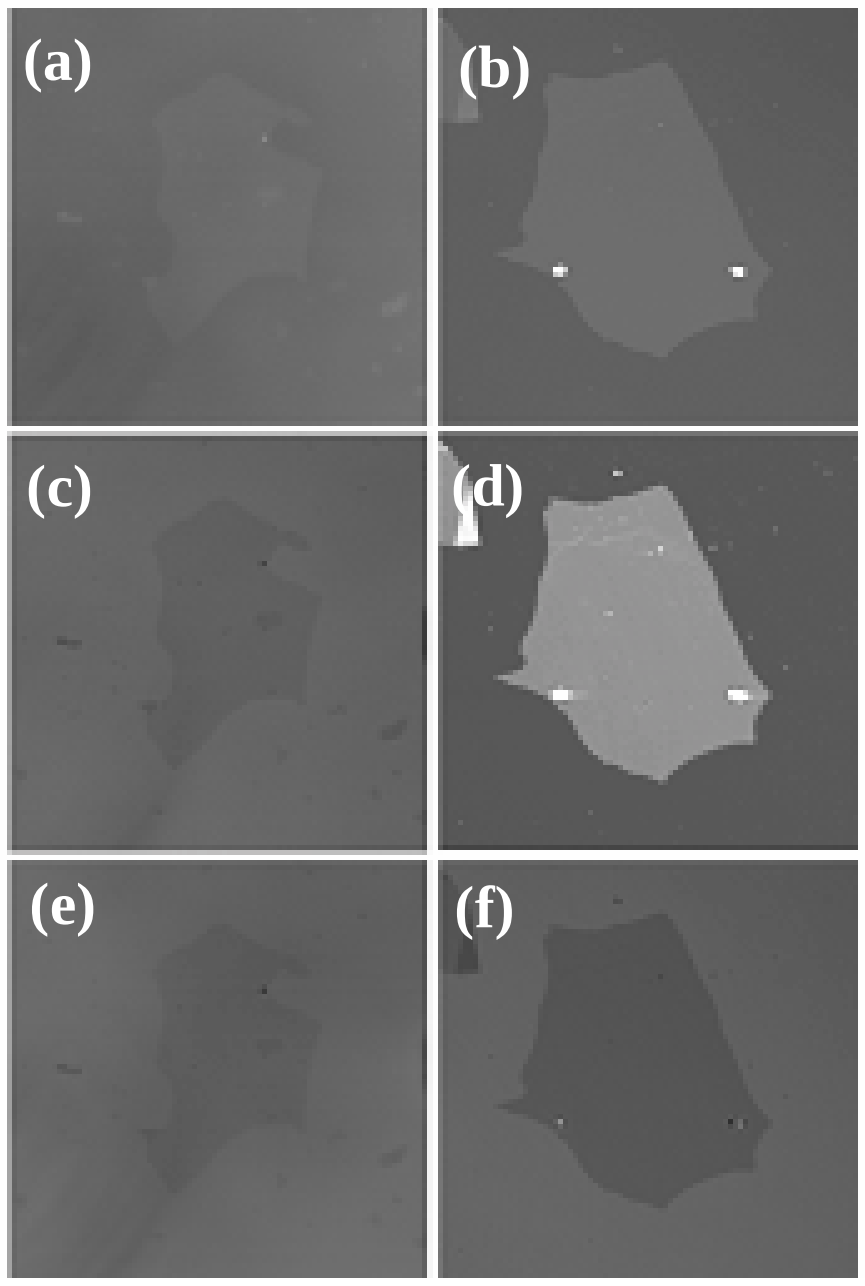

**Figure S-6.** Confocal-microscope images of a graphene oxide single sheet for three different incident wavelengths: (a, b) 488 nm, (c, d) 543 nm, and (e, f) 633 nm, for a 270-nm thick silicon dioxide layer (a, c, e) and a 68-nm thick silicon nitride layer (b, d, f).

Variable wavelength confocal microscopy can give even greater contrast. For a graphene oxide single sheet, a maximum value of contrast higher than 0.8 and minimum value lower than -0.6 were achieved. Sample images are shown in Figure S-8.

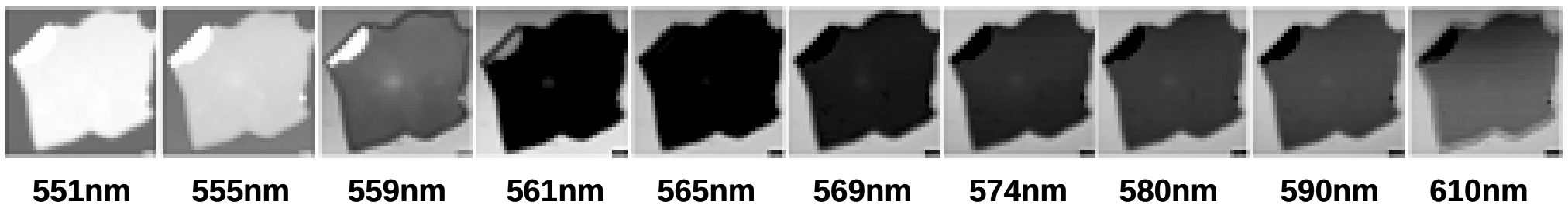

**Figure S-7.** Variable wavelength confocal microscope images of a graphene oxide sheet, for a 67-nm thick silicon nitride layer.

***Silicon nitride grown by low stress and stoichiometric methods***

The optical properties of silicon nitride depend on the growth method used. In figure S-8, the optical properties of two different nitrides are shown along with the calculated contrast, based on their optical properties.

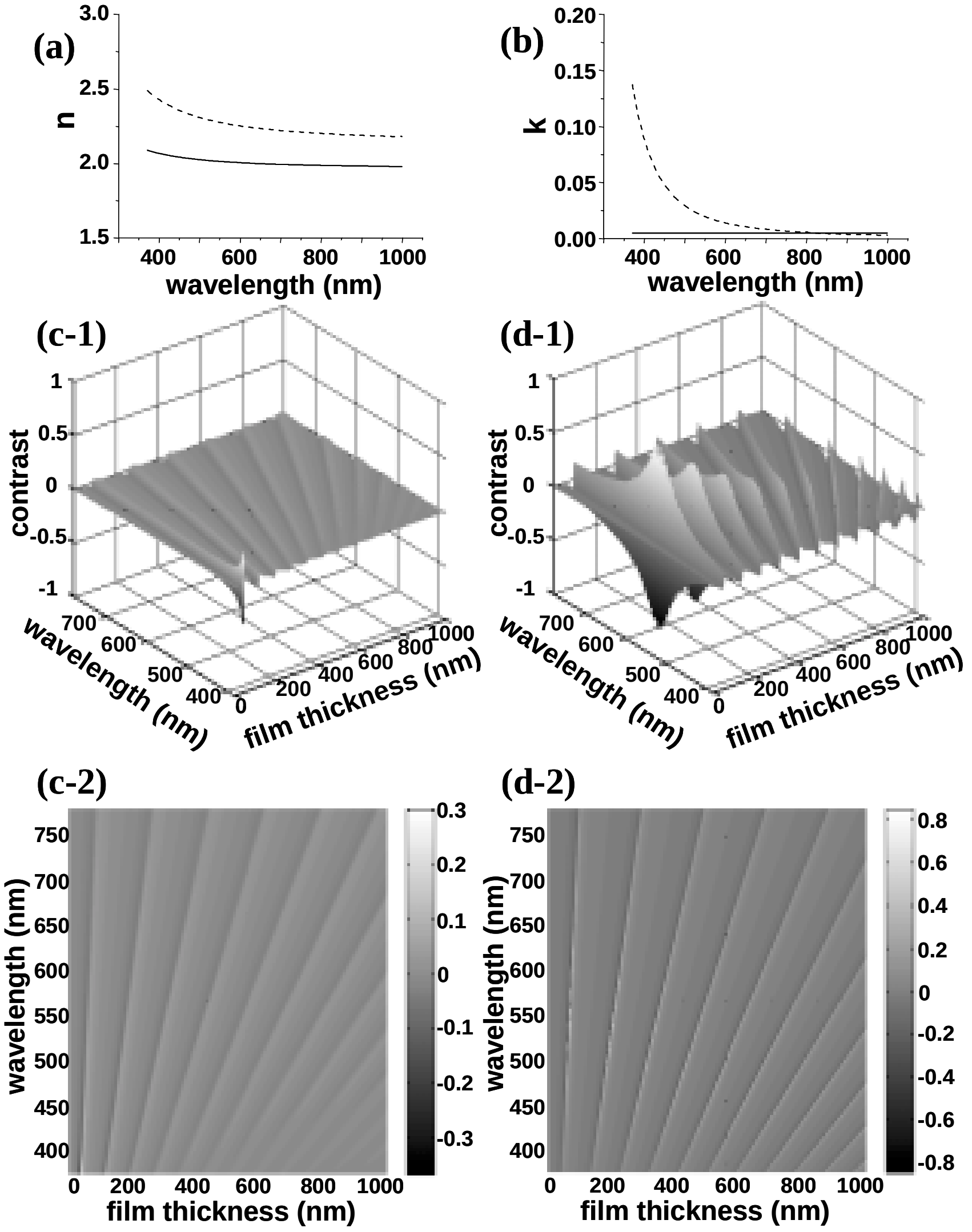

**Figure S-8**. Index of refraction (a) and extinction coefficient (b) of silicon nitride from different growth methods: stoichiometric (solid line) and low stress (dashed line); contrast as a function of wavelength and film thickness for (c-1, c-2) silicon nitride from the low stress growth method and (d-1, d-2) from the stoichiometric growth method.

The calculated contrast and the real images, figure S-9, show that the silicon nitride from the stoichiometric growth method is more effective for obtaining high contrast.

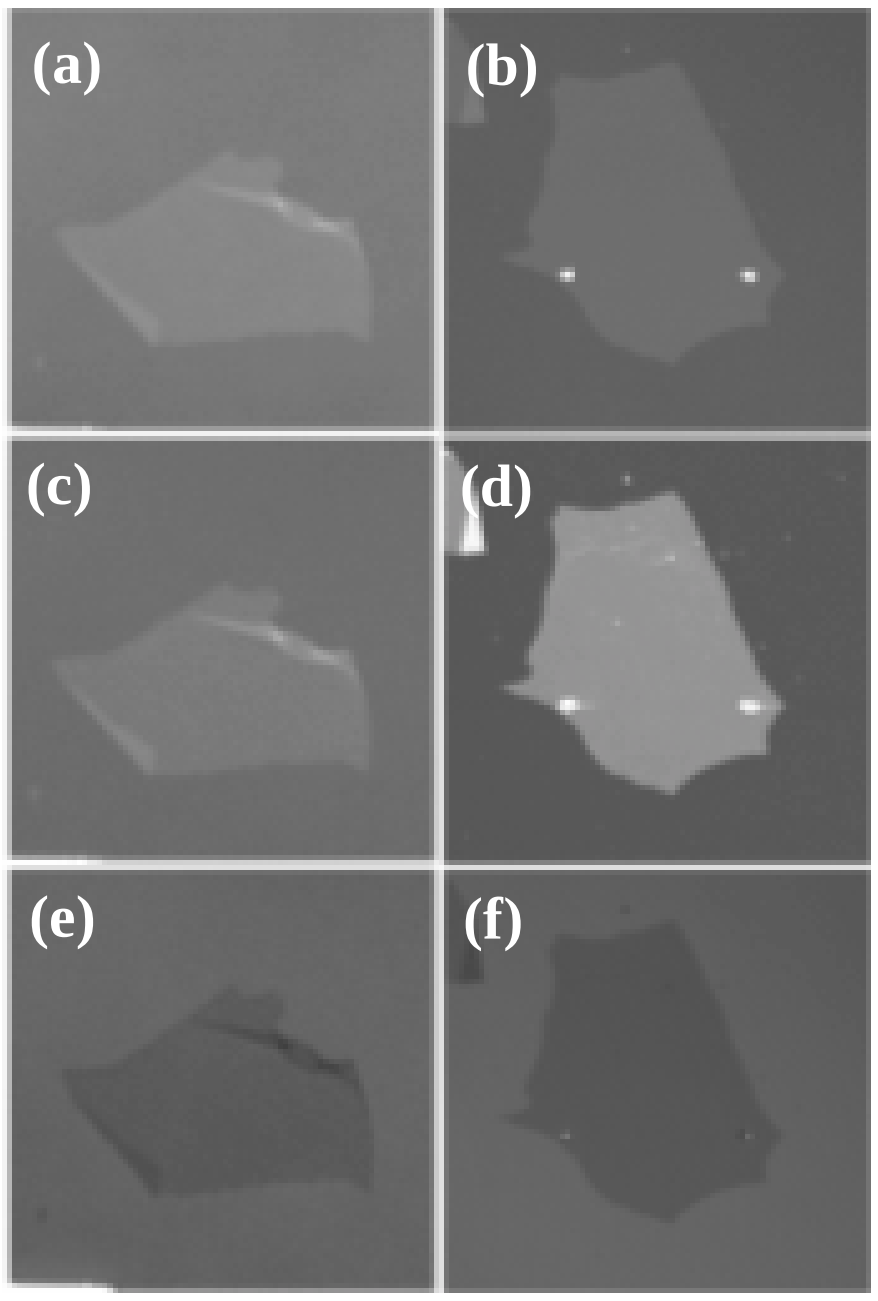

**Figure S-9.** Confocal laser scanning microscope images of a graphene oxide single sheet on 67-nm thick “low-stress” silicon nitride (a, c, e) and on 68-nm thick “stoichiometric” silicon nitride, at three different wavelengths: (a, b) 488 nm, (c, d) 543 nm, and (e, f) 633 nm.

## 4. Uncertainty of optimized optical properties.

***Dielectric layer thickness***

In order to determine the uncertainties in our fitting procedure, we perform additional analysis of the fitting for the wavelength-dependent data shown in figure 6. In figure S-10, the effect of varying the assumed dielectric layer thickness is shown. The mean square error between calculation and experiment is minimized when the thickness is assumed to be 67.4 nm (see figure S-10(c)). Assuming that the uncertainty of the dielectric layer thickness is ±0.2nm, the uncertainties of optimized *n* and *k* are ±0.2 and ±0.1, respectively.

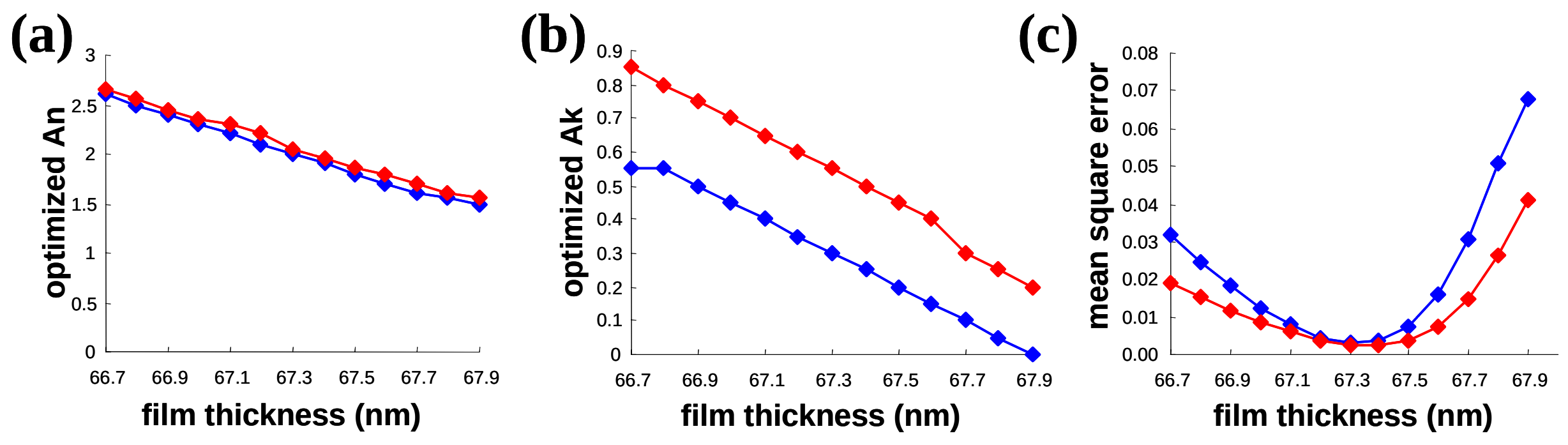

**Figure S-10** Effect of assumed dielectric layer thickness on optimized optical properties and mean square error, before (blue) and after (red) the thermal treatment.

***Graphene-oxide thickness***

The uncertainty of the thickness of the graphene-oxide layer also affects the optimized optical properties. In this case, the mean square error has no minimum along the change of the material thickness. Assuming that the uncertainty of the graphene oxide thickness is ±0.2nm, the uncertainties of optimized *n* and *k* are ±0.2 and ±0.1, respectively.

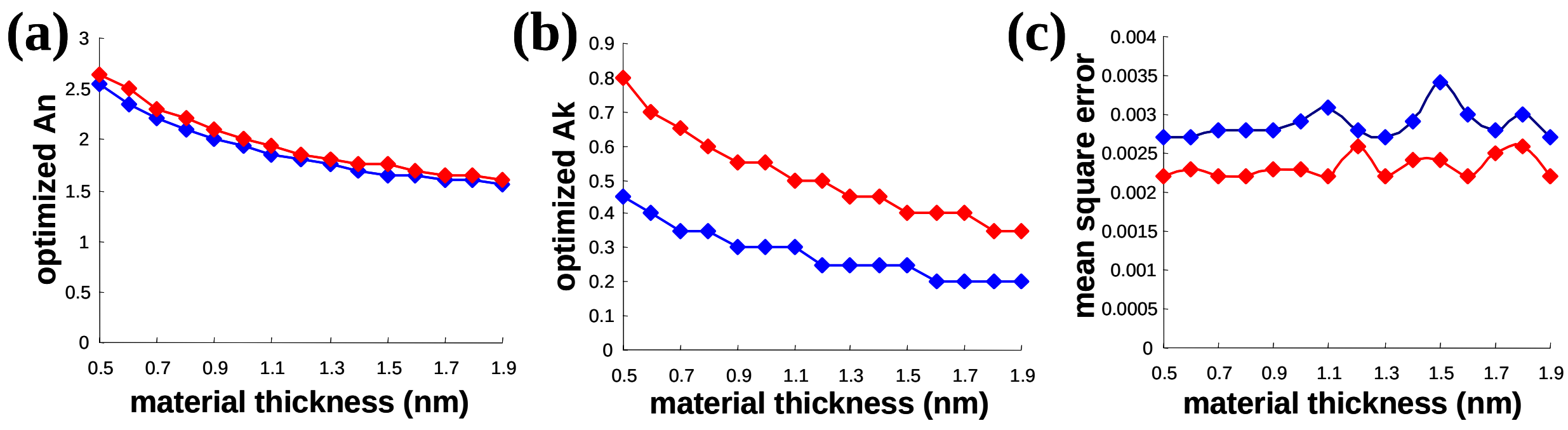

**Figure S-11** Effect of assumed graphene-oxide thickness on optimized optical properties and mean square error: before (blue) and after (red) the thermal treatment.